\documentclass[aps,twocolumn,showpacs,amsmath,amssymb,pre,superscriptaddress, floatfix]{revtex4-1}

\usepackage{graphicx}% Include figure files
\usepackage{dcolumn}% Align table columns on decimal point
\usepackage{bm}% bold math
\usepackage{amssymb}
\usepackage{multirow}

\begin{document}

\title{Disorder-induced regular dynamics in oscillating lattices}

\author{Thomas Wulf}
    \email{Thomas.Wulf@physnet.uni-hamburg.de}
    \affiliation{Zentrum f\"ur Optische Quantentechnologien, Universit\"at Hamburg, Luruper Chaussee 149, 22761 Hamburg, Germany}
\author{Benno Liebchen}
    \affiliation{Zentrum f\"ur Optische Quantentechnologien, Universit\"at Hamburg, Luruper Chaussee 149, 22761 Hamburg, Germany}

\author{Peter Schmelcher}
    \email{Peter.Schmelcher@physnet.uni-hamburg.de}
    \affiliation{Zentrum f\"ur Optische Quantentechnologien, Universit\"at Hamburg, Luruper Chaussee 149, 22761 Hamburg, Germany}
    \affiliation{The Hamburg Centre for Ultrafast Imaging, Universit\"at Hamburg, Luruper Chaussee 149, 22761 Hamburg, Germany} 

\date{\today}


\begin{abstract}

We explore the impact of weak disorder on the dynamics of classical particles in a periodically oscillating lattice.
It is demonstrated that the disorder induces a hopping process from diffusive to regular motion i.e. we observe the counterintuitive phenomenon that disorder leads to regular behaviour.
If the disorder is localized in a finite-sized part of the lattice,
the described hopping causes initially diffusive particles to even accumulate in regular structures
of the corresponding phase space. A hallmark of this accumulation is the emergence of pronounced peaks
in the velocity distribution of particles which should be detectable in state of the art experiments e.g. with cold atoms in optical lattices.

\end{abstract}

\maketitle
%%%%%%%%%%%%%%%%%%%%%%%%%%%%%%%%%%%%%%%

\paragraph*{Introduction.}
-- Time-driven nonequilibrium dynamics 
is a subject of major interest \cite{ Denisov:2006, Gong:2006, Wang:2007, Salger:2009, Haenggi:2009},
covering many different physical systems such as  
colloidal particles exposed to periodically modulated ion chains \cite{Bleil:2007}, particles moving along a filament with a hydrodynamic coupling to the surrounding solvent \cite{Malgaretti:2012}
or cold polar atoms loaded into optical lattices that are driven by periodic phase modulations of the applied laser beam \cite{Renzoni:2006, Renzoni:2010, Renzoni:2012}.
A prototype example for a nonequilibrium phenomenon in driven lattices is the     
celebrated 'ratchet effect' which is the appearance of directed particle motion in the absence 
of biased forces due to a breaking of certain spatiotemporal symmetries \cite{Flach:2000, Quintero:2010}.
While the aforementioned setups focus on globally acting time periodic forces,  
it has been demonstrated recently how spatially varying 
ac-forces introduce a plethora of effects \cite{Reimann:2007, Petri:2010, Liebchen:2011, Petri:2011, Wulf:2012} such as the formation of 
density waves \cite{Petri:2011}, the patterned deposition of particles \cite{Liebchen:2011} or the possibility for conversion processes between diffusive and 
ballistic motion \cite{Petri:2011, Wulf:2012}. 
Moreover, long-range interactions have been shown to lead to dynamical current reversals in the absence of any paramter change \cite{Liebchen:2012}. 
In this work we demonstrate how the combination of disorder and driving in a lattice can lead to the emergence of regular motion 
from an originally chaotic and diffusive ensemble of particles.
Disorder-induced autocorrelations and pronounced changes of the velocity distributions are found with major differences 
occuring for the cases of global versus local disorder.

%%%%%%%%%%%%%%%%%%%%%%%%%%%%%%%%%%%%%%%%%%%%%%%%%%%%%%%%%%%%%%%%%%%%%%%%%%%%%%%%%%%%%%%%%%%%%%%%%%%%%%%%%%%%%%%%%%%%%%%%%%%%%%%%%%%%%%%%%%%%%%%%%%%%%%%%%%%%%%%%%%%%%%%%%%%%%%%%%%%%%%%%%%%%%%%%%
\paragraph*{The driven lattice Hamiltonian.}
--We consider a one-dimensional lattice, consisting of laterally oscillating square barriers of width $l$ and
spacing $L$. Each barrier is labeled by an index $n_b \in \mathbb{Z}$ on which its potential height $V$ depends, i.e. $V=V(n_b)$.     
The classical Hamiltonian for non-interacting particles is thus given by  
\begin{equation}
  H(x,p,t)= \frac{p^2}{2m}+ \sum_{n_b=-\infty}^{\infty} V(n_b) \cdot \Theta (l/2-|x-X_{0,n_b}-d(t)|),
  \label{Hamiltonian}
\end{equation}
where $X_{0,n_b}$ is the equilibrium position of the $n_b$th barrier and $d(t)=A(\cos(\omega t)+\sin(2\omega t))$
is the driving law. By this choice of $d(t)$ we break the relevant symmetries to allow for directed transport \cite{Flach:2000, Quintero:2010}.
We note that the effects and phenomena observed in this work do not rely on the specific shape of the lattice potential.\\
While for a uniform lattice all barriers are indistinguishable with $V(n_b)\equiv V_0$, disordered lattices can be realized by 'perturbing'
some barriers randomly, e.g. by setting $V(n_b)< V_0$ for some $n_b$. 
More precisely, we define a distribution $\sigma: n_b \in \mathbb{Z} \mapsto X \in (0,1)$	%\mathbb{R}| 0<\chi<1$     
which assigns a random number to each barrier and determines its potential height via:
\begin{equation}
 V(n_b)=\left\{\begin{array}{cl} V_0-\eta \sigma(n_b), & \quad \mbox{for} \quad \ \sigma(n_b) \geq \gamma \quad \text{and} \quad |n_b|<D
 \\ V_0, &  \quad \mbox{else} \end{array}\right. 
\label{Potential}
\end{equation}
where $\gamma$ describes the relative amount of perturbed barriers, $\eta$ controls the perturbation strength and $D$ accounts for the extension of the disorder region.
For $D=0$ no barrier is perturbed and the lattice is said to be uniform. For $D\rightarrow \infty$ the lattice is globally disordered and for a finite nonzero $D$ we obtain localized disorder.
Note that by this procedure, both the locations of the perturbed barriers as well as the perturbation strengths are chosen randomly and are not correlated with one another.\\
The focus of this work is on the regime of weak disorder which is accounted for by choosing $\gamma=0.9$, i.e. for approximately 90 $\%$ of the barriers we set $V(n_b) = V_0$.
The remaining parameters are fixed to: $m=V_0=\omega=1.0$, $L=5.0$, $\eta =0.9$, $A\approx 0.57$ corresponding to a maximal barrier displacement of max$|d(t)|=1.0$.
Before we discuss the impact of disorder let us briefly account for some of the main aspects of the dynamics in the uniformly driven lattice.

\begin{figure}[htbp]
\centering
\includegraphics[width=1.0\columnwidth]{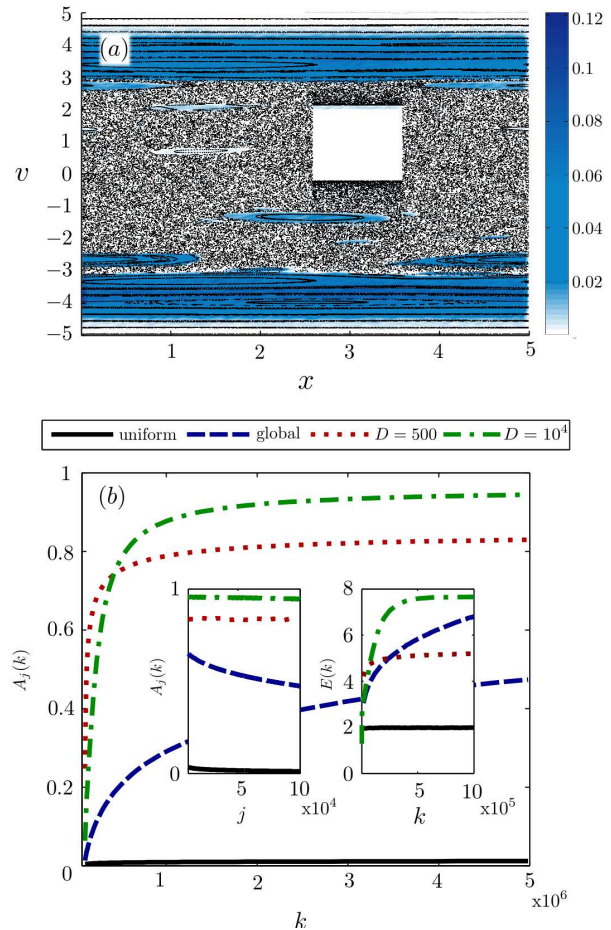}
\caption{\label{fig:autocorr}(a) In black: Stroboscopic PSS of a driven lattice without disorder. For parameters see main text. Coloring: Phase space occupation of an ensemble at $t=10^6$ for localized disorder ($D=10^4$).
  (b) Dependence of ensemble averaged VACFs $ A_j(k)$ on the number of particle-barrier collisions $k$ for $j=10^5$.
  Setups: uniform lattice (black straight line); global disorder (blue dashed line); localized disorder for $D=500$ (red dotted line) and $D=10^4$ (green dashed and dotted line) respectively. 
  The left inset shows $A_j(k)$ as a function of $j$ for $k=5\cdot 10^6$. The right inset shows the mean kinetic energy of particle ensembles as a function of $k$.}
\end{figure}

%%%%%%%%%%%%%%%%%%%%%%%%%%%%%%%%%%%%%%%%%%%%%%%%%%%%%%%%%%%%%%%%%%%%%%%%%%%%%%%%%%%%%%%%%%%%%%%%%%%%%%%%%%%%%%%%%%%%%%%%%%%%%%%%%%%%%%%%%%%%%%%%%%%%%%%%%%%%%%%%%%%%%%%%%%%%%%%%%%%%%%%%%%%%%%%%%
\paragraph*{Dynamics in the unperturbed system.}
--The Hamiltonian given in eqs. (\ref{Hamiltonian}) and (\ref{Potential}) contains the uniformly driven lattice with $V(n_b)\equiv V_0$ for all $n_b$ by setting $D=0$. 
As described in detail in \cite{Wulf:2012, Petri:2010}
such a setup features a mixed phase space containing a 'chaotic sea' at low kinetic energies which is bounded by invariant curves and shows regular islands 
embedded in the chaotic sea. The stroboscopic Poincar\'{e} surfaces of section (PSS), where at times $t=2\pi\cdot n$ with $n \in \mathbb{N}$ velocity and position $x$ modulus the barrier distance $L$ are recorded,
is shown in Fig. \ref{fig:autocorr} (a) for the parameters as mentioned above.
Correspondingly, there are two distinguished types of motion apparent: Either a trajectory is located within the chaotic sea 
or it is located within any of the regular structures.
Opposite to the diffusive motion in the chaotic sea the regular motion for islands above or below $v=0$ preserves the sign 
of the particle velocity
i.e. $v(t)\cdot v(t+\Delta t)>0$ for any $t$ and $\Delta t$
and corresponds to ballistic motion.
Note that the dynamics of a particle in the chaotic sea typically contains also phases of 'ballistic flights' of arbitrary but finite length, i.e. a 
diffusive particle can undergo ballistic motion for a limited amount of time. The lengths of these ballistic 
flights obey a power law distribution \cite{Petri:2010}.

%%%%%%%%%%%%%%%%%%%%%%%%%%%%%%%%%%%%%%%%%%%%%%%%%%%%%%%%%%%%%%%%%%%%%%%%%%%%%%%%%%%%%%%%%%%%%%%%%%%%%%%%%%%%%%%%%%%%%%%%%%%%%%%%%%%%%%%%%%%%%%%%%%%%%%%%%%%%%%%%%%%%%%%%%%%%%%%%%%%%%%%%%%%%%%%%%
% \section{Dynamics in the disordered system}
% \label{section4}
\paragraph*{Dynamics in the disordered system.}
We now demonstrate how the presence of disorder influences the previously described dynamics in the driven lattice. 
A suitable observable to distinguish between diffusive and ballistic motion is the autocorrelation function, which as we shall see below is altered significantly 
by the inclusion of disorder.
Velocity distributions represent a second valuable tool to analyze the impact of disorder and are discussed afterwards.

\begin{figure}[htbp]
\centering
\includegraphics[width=1.0\columnwidth]{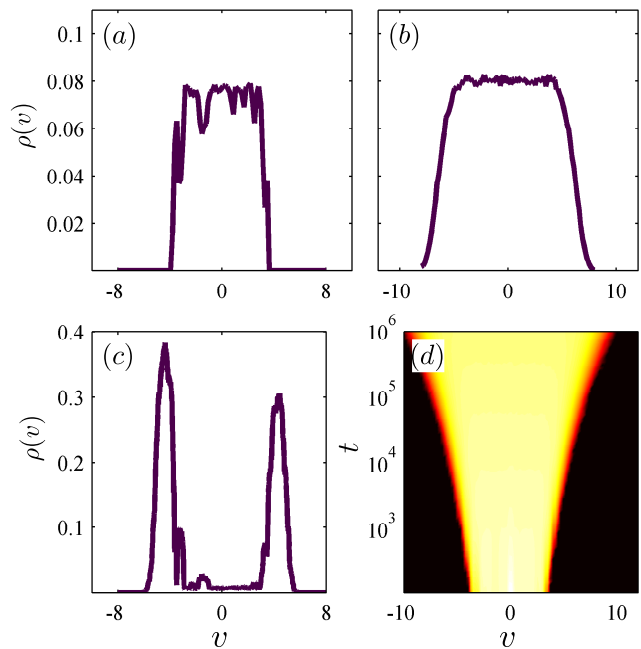}
\caption{Velocity distributions $\rho(v)$ at $t=10^6$ for uniform driving (a), global disorder (b) and localized disorder with $D=10^4$ (c). (d) Time evolution of $\rho(v)$ for the same setup as in (b).
The color scaling is logarithmic for better visibility.\label{fig:velo_dists}}
\end{figure}

\begin{figure}[htbp]
\centering
\includegraphics[width=1.0\columnwidth]{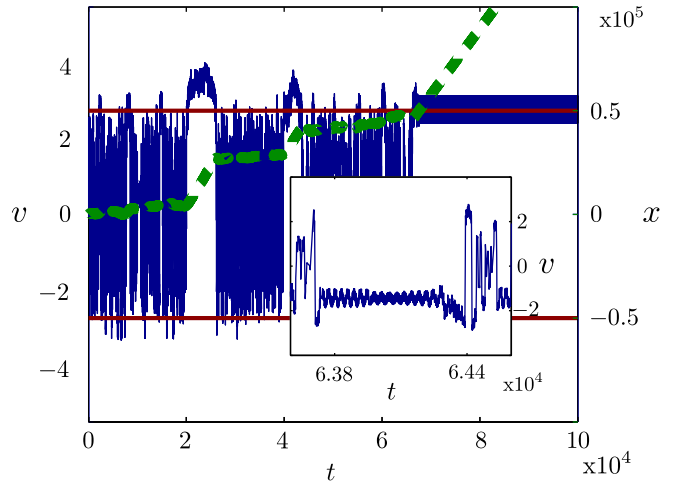}
\caption{Position (right ordinate, green dashed line) and velocity (left ordinate, blue line) as a function of time for a single trajectory. The horizontal red lines are at $x=\pm x_D=\pm 5\cdot 10^4$ and delimit the region of disorder. 
The inset is a zoom into $v(t)$, indicating the entering and exiting processes
of a quasi periodic orbit.\label{fig:traj}}
\end{figure}

%%%%%%%%%%%%%%%%%%%%%%%%%%%%%%%%%%%%%%%%%%%%%%%%%%%%%%%%%%%%%%%%%%%%%%%%%%%%%%%%%%%%%%%%%%%%%%%%%%%%%%%%%%%%%%%%%%%%%%%%%%%%%%%%%%%%%%%%%%%%%%%%%%%%%%%%%%%%%%%%%%%%%%%%%%%%%%%%%%%%%%%%%%%%%%%%%
% \subsection{Disorder-induced autocorrelations}
% \label{section4.1}
The velocity autocorrelation function (VACF):
\begin{equation}
 A_j(k)=\frac{1}{k-j} \sum_{i=1}^{k-j} \frac{2v_iv_{i+j}}{v_i^2+v_{i+j}^2}, \qquad k>j
\label{eq1}
\end{equation}
relates the velocity of a particle after its $i$th collision $v_i$ with its velocity after the $(i+j)$th collision with one of the barriers and the normalization ensures that $-1\leq A_j(k) \leq 1$.
On the one hand, if a particle moves diffusively through the lattice, its velocity is allowed to switch its sign and different terms of the sum can cancel each other. 
On the other hand, if a particle moves ballistically, $v_i$ and $v_{i+j}$ have the same sign and thus the terms in eq. (\ref{eq1}) add up. 
Hence, $A_j(k)$ is a useful quantity to distinguish between ballistic- and diffusive motion in the lattice.
As a starting point we calculate the ensemble averaged VACF numerically for the uniform setup, i.e. $D=0$ in eq. (\ref{Potential}), for $j=10^5$. 
For this purpose we propagate in total $10^6$ trajectories for $100$ different disorder realizations, i.e. for $100$ different random distributions $\sigma(n_b)$.
The initial conditions are $x=0$ and a randomly chosen velocity with $-0.1<v_0<0.1$ in order to ensure that all trajectories are initially located within the chaotic
sea of the unperturbed system. 
The resulting VACF is shown in fig. \ref{fig:autocorr} (b) and can be seen to evolve with increasing collision number $k$
to a small nonzero value. This is a consequence of the directed transport induced by 
the bichromatic driving which is accompanied by an asymmetric chaotic sea with respect to $v=0$ implying a drift of the diffusive chaotic motion (see \cite{Wulf:2012} for details).\\
In the presence of global disorder, i.e. $D\rightarrow \infty$ in eq. (\ref{Potential}), we observe a much stronger increase of the VACF which is even more pronounced
for the two setups with localized disorder as shown for $D=500$ and $D=10^4$. Note that while the VACF appears to saturate for both locally
disordered setups, there is no sign of saturation for the globally disordered lattice.  
To ensure, that the different behaviour in the cases of localized and global disorder is not an artifact of the particular choice of the delay $j$,
we show additionally the VACFs for fixed $k=5\cdot 10^6$ as a function of $j$ in the left inset of fig. \ref{fig:autocorr} (b). Apparently, the increase of $A_j(k)$ does not
require any specific choice of the delay $j$ but is present over the entire investigated parameter regime.
This also shows that for localized disorder $ A_j(k)$ appears to be independent of $j$ whereas for the 
globally disordered setup it decays with increasing delay $j$. Finally, we remark that in the absence of disorder, i.e. for the uniform setup, it is well known that 
correlations decay exponentially with $j$.\\
%%%%%%%%%%%%%%%%%%%%%%%%%%%%%%%%%%%%%%%%%%%%%%%%%%%%%%%%%%%%%%%%%%%%%%%%%%%%%%%%%%%%%%%%%%%%%%%%%%%%%%%%%%%%%%%%%%%%%%%%%%%%%%%%%%%%%%%%%%%%%%%%%%%%%%%%%%%%%%%%%%%%%%%%%%%%%%%%%%%%%%%%%%%%%%%%%
% \subsection{Impact of disorder on velocity distributions}
% \label{section4.2}
Before we analyze the observed behaviour of the autocorrelations in the presence of disorder in more detail, it is instructive
to investigate how the velocity distributions of particle ensembles are influenced by the disorder.
These distributions $\rho(v)$ were obtained by propagating particle ensembles in different setups containing global-, localized- or no disorder. 
The initial conditions are chosen as before ensuring that no particles are initially within regular structures of the unperturbed system.
After a simulation time of $t=10^6$ the velocity of each particle is recorded yielding the distributions $\rho(v)$ as shown in figs. \ref{fig:velo_dists} (a)-(c).
For the uniformly driven lattice (fig. \ref{fig:velo_dists} (a)) we observe that all particles are confined within a velocity interval $-5\lesssim v \lesssim +5$ and
a further inspection reveals that after a certain transient time of $t\approx10^4$, $\rho(v)$ becomes stationary.
This behaviour is easily understood by inspecting the phase space of the unperturbed system (fig. \ref{fig:autocorr} (a)): 
All particles are initially located within the chaotic sea and are occupying it uniformly after a transient time. Because the chaotic sea is bounded by invariant curves the velocity distribution is bounded as well. 
The apparent dips in $\rho(v)$, e.g. at $v\approx -2$ are caused by regular islands, i.e. 
parts of the phase space that are prohibited for chaotic trajectories and thus lead to a reduced density of chaotic trajectories
in these velocity intervals.\\
For the globally disordered lattice (cf. fig. \ref{fig:velo_dists} (b)) the pronounced dips as seen in fig. 
\ref{fig:velo_dists} (a) cease to exist. Moreover the time evolution of $\rho(v)$ (fig. \ref{fig:velo_dists} (d)) indicates, that no stationary distribution is reached. 
On the contrary, $\rho(v)$ spreads to higher and higher velocities as time proceeds.   
In some sense, the effect of a broadening of the velocity distribution appears to be reversed for the setup of localized disorder with $D=10^4$ (fig. \ref{fig:velo_dists} (c)) 
where we observe two pronounced peaks accompanied by smaller ones. 
Similar to the uniformly driven lattice, $\rho(v)$ reaches a stationary form after a transient time of approximately $t\approx 2\cdot 10^5$.
The claim that $\rho(v)$ becomes stationary for both no disorder as well as for localized disorder while it does not for global disorder is 
supported by investigating the mean kinetic energy of the particle ensemble $\langle E \rangle$ 
as a function of the number of collisions $k$ as shown in the right inset of fig. \ref{fig:autocorr} (b). 
While it saturates for the cases of localized- and no disorder, there are no signs of saturation in the presence of global disorder.
Besides the fact that the peaks in $\rho(v)$ are slightly less prominent compared to the background, the setup for $D=500$, which 
was considered previously, reveals no qualitatively different features in comparison with the case $D=10^4$.
%%%%%%%%%%%%%%%%%%%%%%%%%%%%%%%%%%%%%%%%%%%%%%%%%%%%%%%%%%%%%%%%%%%%%%%%%%%%%%%%%%%%%%%%%%%%%%%%%%%%%%%%%%%%%%%%%%%%%%%%%%%%%%%%%%%%%%%%%%%%%%%%%%%%%%%%%%%%%%%%%%%%%%%%%%%%%%%%%%%%%%%%%%%%%%%%%
% \section{Particle accumulation in regular islands}
% \label{section5}
\paragraph*{Particle accumulation in regular islands.}
Let us now explore the physical mechanism behind the observed impact of disorder on the velocity distributions as well as on the VACFs. 
We first discuss the case of global disorder and explain the differences to local disorder afterwards.\\
%%%%%%%%%%%%%%%%%%%%%%%%%%%%%%%%%
% \subsection{Global disorder}
% \label{section5.1}
As argued above, the choice $\gamma=0.1$ ensures that most barriers remain unperturbed. 
Hence, we can expect to
gain insight into the dynamical processes of the disordered system by performing projections onto the phase space of the unperturbed system (PSUS).
In the disorder-free lattice trajectories belong either to the chaotic sea or to one of the regular islands for all times.
In a system with weak disorder the scattering of a particle off a barrier modified due to the disorder can be interpreted 
as a hopping process in the PSUS.
The crucial observation is now, that the corresponding shift in phase space
may transport the particle from the chaotic sea onto a regular island of the PSUS, in which it is trapped until a further collision with a disorder
barrier occurs and the particle may either reenter the chaotic sea or remain within the island. 
Such an island entering event is shown exemplarily in the inset of fig. \ref{fig:traj} where the velocity as a function of time is shown for a single trajectory. 
While the particle moves diffusively at early- as well as at late times, it follows a quasiperiodic orbit of a regular island and thus moves ballistically for intermediate times.
The fact that the velocity distributions for the globally disordered lattice (fig. \ref{fig:velo_dists} (b)) features neither pronounced dips nor peaks, suggests 
that at the time when $\rho(v)$ is recorded both the chaotic- as well as the regular islands of the PSUS are populated uniformly.
At the same time particles can become faster than in the uniformly driven lattice by entering regular spanning curves at velocities above the chaotic sea. This 
leads to the spreading of $\rho(v)$ to increasingly higher velocities (fig. \ref{fig:velo_dists} (d)) as well as to the increase of the mean kinetic energy (right inset of fig. \ref{fig:autocorr} (b))
and is reminicent of Fermi acceleration \cite{Eichler:1987} as occuring e.g. in driven billards \cite{Lenz:2008}.\\
%%%%%%%%%%%%%%%%%%%%%%%%%%%%%%%%
% \subsection{Local disorder}
% \label{section5.2}
We now address the question why localized disorder both leads to peaked velocity distributions and increases the VACF.
As argued above a value of the VACF close to one, as observed at large $k$ for the setups with 
localized disorder, serves as an indicator that particles predominantly undergo regular instead of diffusive motion. 
Indeed, fig. \ref{fig:autocorr} (a) which shows the occupation of the phase space on top of the PSUS for the ensemble at $t=10^6$ for localized disorder with $D=10^4$, 
reveals that peaks in $\rho(v)$ as shown in fig. \ref{fig:velo_dists} (c) are caused by 
an accumulation of particles within regular structures of the phase space of the uniform part of the lattice.
More precisely, the two dominant peaks correspond to particles
on invariant spanning curves, while the smaller peaks are related to particles in regular islands of the PSUS. In the following we identify two 
distinct mechanisms which both contribute to the observed accumulation in regular structures.\\
%%%%%%%%%%%%%%%%%%%%%%%%%%%%%%%%%%%%%%%%%%%%%%%%%%%
\textit{1. Multiple entering of disorder region:}
Apparently, the above described conversion processes from regular to chaotic motion or vice versa can only occur for particles which are still within the 
disorder region, that is if $|x|<x_D \equiv D \cdot L$. Once the particle passes $x_D$ we have to distinguish two different scenarios: 
Either the particle passes $x_D$ while being in a regular structure of the PSUS or it is located in the chaotic sea of the PSUS.  
In the former case, the particle does not 
reenter the disordered region and thus cannot leave its regular 'host' because no further collisions with perturbed barriers occur. 
Such an event is shown in fig.\ref{fig:traj}: The considered particle switches between diffusive and ballistic phases until it eventually reaches $x_D$ within a regular structure of the PSUS. Afterward 
the particle remains in the corresponding quasiperiodic orbit.
If on the other hand the particle crosses the edge of the disorder region at $x_D$ while being within the chaotic sea of the PSUS it can cross this edge several times 
since the diffusive chaotic motion can transport the particle back to the position $x_D$. Accordingly, it 
has several chances of being transformed into a ballistic particle by means of a collision with a perturbed barrier.
Thus, the possibility to cross the edge at $x_D$ several times as long as the particle remains chaotic is one of the reasons why the majority of particles become regular in the lattices with localized finite disorder.\\
%%%%%%%%%%%%%%%%%%%%%%%%%%%%%%%%%%%%%%%%%%%%%%%   
\textit{2. Separation of length scales:}
Following the previous arguments, the edges of the disorder region at $x=\pm x_D$ are expected to play a prominent role for the particle accumulation in the regular parts of the PSUS. This is supported by the observation that 
$A_j (k)$ is larger for $D=500$ than for $D=10^4$ for small $k$ because it takes less collisions to reach $x=\pm x_D$.
However, it is also apparent that $A_j (k)$ for $D=10^4$ outgrows the VACF for $D=500$ for larger values of $k$.
Further inspection reveals, that this is due to a separation of length scales. Consider a particle located near the center of the disorder region.
As fig. \ref{fig:traj} illustrates, such a particle undergoes both diffusive as well as ballistic phases, while the average
distance covered within a diffusive phase, denoted by $l_{\text{d}}$, is smaller than it is in a ballistic phase, because the former includes frequent changes in the sign of the velocity 
while the latter one does not. Consequently, the probability that a particle
reaches the edge of the disorder region $x_D$ within a ballistic phase can be expected to be higher than it is for a diffusive phase. 
Hence this separation of length scales leads to an accumulation of particles into regular structures of the PSUS iff $x_D>>l_{\text{d}}$
which is not the case for $D=500$ and hence this mechanism is partly suppressed in this setup, yielding the lower value of the VACF compared to 
the $D=10^4$ setup. 
However, once $D$ is chosen sufficiently large to allow for this mechanism a further increase of $D$ should not lead to a further enhancement of $A_j(k)$. This has been confirmed numerically
revealing that indeed for $D\gtrsim10^4$ the value of $A_j(k)$ for large $k$ becomes to a good approximation independent of $D$.   

%%%%%%%%%%%%%%%%%%%%%%%%%%%%%%%%%%%%%%%%%%%%%%%%%%%%%%%%%%%%%%%%%%%%%%%%%%%%%%%%%%%%%%%%%%%%%%%%%%%%%%%%%%%%%%%%%%%%%%%%%%%%%%%%%%%%%%%%%%%%%%%%%%%%%%%%%%%%%%%%%%%%%%%%%%%%%%%%%%%%%%%%%%%%%%%%%
% \section{Conclusion}   
% \label{section6}
\paragraph*{Conclusions.}

We have investigated the impact of disorder on the nonequilibrium dynamics of classical particles in a one-dimensional
driven lattice. The disorder causes initially diffusive chaotic particles to enter regular regimes of the phase space.
As hallmarks of these processes we observed both the emergence of pronounced peaks in the velocity distribution of
particle ensembles as well as synchronized particle motion if the disorder is localized in a finite region of the lattice.
Since none of the presented phenomena requires any fine tuning of the system parameters and rely solely on the existence
of a mixed phase space, we believe that our findings are of relevance in understanding the dynamics in different experimental
setups such as cold atoms in optical lattices \cite{Renzoni:2006, Renzoni:2012, Renzoni:2008}. In these experiments counter 
propagating laser beams which are periodically phase modulated provide a laterally driven quasi 1D lattice potential.
By employing a large detuning between the laser field and the inner atomic transition, the Hamiltonian regime can be reached
and finally, disorder could be introduced by superimposing an optical speckle field as was done e.g. in \cite{White:2009}.
Typically in experiments with cold thermal atoms \cite{Renzoni:2006, Renzoni:2012, Renzoni:2008}, the atoms momentum
distribution is recorded after they are exposed to the lattice potential for approx. $10$ms \cite{Renzoni:2008} corresponding
to approx. $10^3$ periods of the lattice oscillations for $\omega=100kHz$. Our results indicate that these time scales
are sufficient to obtain the corresponding peaks in the particles velocity distribution and we thus believe that disorder-induced
regular behaviour could be observed specifically in cold atom, but also in other, setups.


\begin{thebibliography}{9}

% \bibitem{Klitzing:1980} K.v. Klitzing, G. Dorda and M. Pepper, Phys. Rev. Lett. {\bf 45}, 494 (1980)							%% original hall effects
% \bibitem{Evers:2008} F. Evers, A. D. Mirlin, Rev. Mod. Phys. {\bf 80}, 1355–1417 (2008) 								%% Review on anderson loc
% \bibitem{Billy:2008} J. Billy et al., Nature {\bf 453}, 7197:891-894 (2008)										%% anderson loc
% \bibitem{Roati:2008} G. Roati et al., Nature {\bf 453}, 7197:895-898 (2008)										%% anderson loc


\bibitem{Salger:2009} T. Salger et al., Science {\bf 326}, 1241 (2009)											%% quantum ratchet experiment
\bibitem{Haenggi:2009} P. H\"anggi., F. Marchesoni, Rev. Mod. Phys. {\bf 81}, 387 (2009)								%% Review brownian motors.
\bibitem{Denisov:2006} S. Denisov, S. Flach and P. H\"{a}nggi, Europhys. Lett. {\bf 4}, 588 (2006).							%% Ratchet
\bibitem{Gong:2006} J. Gong and P. Brumer, Phys. Rev. Lett.  {\bf 97}, 240602 (2006).									%% Ratchet
\bibitem{Wang:2007} L. Wang, G. Benenti, G. Casati and B. Li Phys, Rev. Lett. {\bf 99}, 244101 (2007).							%% Ratchet


\bibitem{Bleil:2007} S. Bleil, P. Reimann and C. Bechinger, Phys. Rev. E {\bf 75}, 031117 (2007)							%% directed brownian motion with colloides
\bibitem{Malgaretti:2012} P. Malgaretti, I. Pagonabarraga and D. Frenkel, Phys. Rev. Lett. {\bf 109}, 168101 (2012)					%% running faster together...

% \bibitem{Lewenstein:2007} M. Lewenstein et al., Adv. Phys. {\bf 56}, 243 (2007)										%% quantum disorder report
% \bibitem{White:2009} M. White et al., Phys. Rev. Lett. {\bf 102}, 055301 (2009)										%% quantum disorder 

% \bibitem{Ponomarev:2009} A. V. Ponomarev, S. Denisov and P. H\"{a}nggi, Phys. Rev. Lett. {\bf 102}, 230601 (2009).					%% Quantum motor
% \bibitem{Hauke:2012} P. Hauke et al., Phys. Rev. Lett. {\bf 109}, 145301 (2012).									%% shaken optical lattice
%\bibitem{Zenesini:2009} A. Zenesini, H. Lignier, D. Ciampini, O. Morsch and E. Arimondo, Phys. Rev. Lett. {\bf 102}, 100403 (2009).			%% shaken optical lattice


\bibitem{Renzoni:2006} R. Gommers, S. Denisov and F. Renzoni, Phys. Rev. Lett. {\bf 96}, 240604 (2006).							%% Ratchet
\bibitem{Renzoni:2010} P. Phoonthong, P. Douglas, A. Wickenbrock and F. Renzoni, Phys. Rev. A {\bf 82}, 013406 (2010).					%% Ratchet
\bibitem{Renzoni:2012} A. Wickenbrock, P.C. Holz, N.A. Abdul Wahab, P. Phoonthong, D. Cubero and F. Renzoni, Phys. Rev. Lett. {\bf 108}, 020603 (2012).	%% Ratchet

\bibitem{Flach:2000} S. Flach, O. Yevtushenko, and Y. Zolotaryuk, Phys. Rev. Lett. {\bf 84}, 2358 (2000).						%% FLACH DIRECTED CURRENTS due to symmetry breaking
\bibitem{Quintero:2010} N. R. Quintero, J. A. Cuesta and R. Alvarez-Nodarse, Phys. Rev. E. {\bf 81}, 030102 (2010)				%% QUINTERO analysis of dcs


\bibitem{Reimann:2007} P. Reimann and M. Evstigneev, Europhys. Lett. {\bf 78}, 50004 (2007).								%% Reimann phase mod. driving.
\bibitem{Petri:2010} C. Petri, F. Lenz, F. K. Diakonos and P. Schmelcher, Phys. Rev. E {\bf 81}, 046219 (2010).						%% PETRI DIR.TRANS.IN PHASE-MOD.DRIVEN LATTICES
\bibitem{Liebchen:2011} B. Liebchen, C. Petri, F. Lenz and P. Schmelcher, Europhys. Lett. {\bf 94}, 40001 (2011).					%% Benno	Patterned deposition
\bibitem{Petri:2011} C. Petri, F. Lenz, B. Liebchen, F. K. Diakonos and P. Schmelcher, Europhys. Lett. {\bf 95}, 30005 (2011).				%% PETRI DENSITY WAVES
\bibitem{Wulf:2012} T. Wulf, C. Petri, B. Liebchen and P. Schmelcher, Phys. Rev. E {\bf 86}, 016201 (2012)						%% Mein Paper
\bibitem{Liebchen:2012} B. Liebchen, F. K. Diakonos and P. Schmelcher, New J. of Phys. {\bf 14}, 103032 (2012).					%% Benno	Current Reversals

\bibitem{White:2009} M. White et al., Phys. Rev. Lett. {\bf 102}, 055301 (2009)										%% quantum disorder 

%\bibitem{Fermi:1949} E. Fermi, Phys. Rev. {\bf 75}, 1169 (1949)												%% Fermi acceleration
\bibitem{Eichler:1987} R. Blandford and D. Eichler, Phys. Rep. {\bf 154}, 1 (1987)  									%% Fermi accelerataion review
\bibitem{Lenz:2008} F. Lenz, F.K. Diakonos and P. Schmelcher, Phys. Rev. Lett. {\bf 100}, 240604 (2008).							%% Fermi acceleration billard


\bibitem{Renzoni:2008} R. Gommers, V. Lebedev, M. Brown and F. Renzoni, Phys. Rev. Lett. {\bf 100}, 040603 (2008).					%% Ratchet exp.

\end{thebibliography}
\end{document}